# Users Approach on Providing Feedback for Smart Home Devices

Santhosh Pogaku

University of Cincinnati, Pogakush@mail.uc.edu

## 1. Abstract :

Smart home technology is part of our everyday lives, and this technology is fast-evolving compared to other technologies. The user's feedback is gathered in this paper by conducting expert interviews on how collecting the feedback from the smart home devices will be helpful to improve the devices. We are yet to know about the feedback system of the smart home devices and how provided feedback will support increasing the devices' requirements. Today, we present our analysis from our exploratory interview method with the student of a certain group, and we try to study the attitude of providing feedback. The results suggested that the users are ready to give their feedback very actively to better their usage as every user has their own needs to fulfill.

## 2. Introduction :

The current market of smart homes projects a global revenue of 85 Billion USD, and the usage is read to be around 10% worldwide[1]. Smart home technology combines automation interface, monitors, and sensors [2]. Smart home devices will tend to increase the way of life of humanity and the comfort of many people. The looks are deceiving as smart homes aren't taking the user needs, and necessities into accountability [3] as several critical challenges related to innovative home technology are yet to be addressed for smart home development [4]. Smart home technologies are not very often in the way they should be used, even with specific settings. A clear view of the user's usage of the smart home devices is still missing. Specific methods such as interviewing certain people will not help as innovative home technology evolves [5]. User feedback plays a vital role in getting the desired outcome and helps understand the user needs and what they want to use the smart home devices in the desired way to fill the usage gap [4]. After exploring the internet, user feedback hasn't been explored in smart home technology to date. We can think of essential features of smart home devices as UI, Networking, and Interface that make us re-visits or re-think the current user feedback approaches. The exploration has not changed the current feedback methods for smart home parameters or requirements. We can take an example of the users who are using the smart home technology and genuinely how much they are providing feedback to the devices during their daily routines regarding their personal space as it's a sensitive area. Innovative home solutions are interconnected all the time, and the ecosystem has different types of devices, and a few of them are sensors where no human intervention is needed. The smart home products are different than regular products. These products are present in the living space of humankind. Yet, these devices even take care of us, and it brings a brand-new facet to such kind of feedback. Hence, from the industry and researchers' perspective, it is evident that there is a need for research to build excellent feedback tools or models[6]. The long-term research goal would be to build group-based feedback solutions for smart devices. This paper will use exploratory research methodology and showcase the result from an exploratory interview study to analyze the users' feedback on smart home technology. At the beginning of the design, the software and the engineering requirements will define the user feedback [7]. The feedback designed is very meaningful and designed based on the existing experiences and would be helpful for future software development [8]. There is a term named push feedback [7], where users can give feedback on bot linguistic and non-linguistic formats. In the past, research has been conducted, and the toll support user feedback has been used, but it has dedicated settings[9] [10]. Many studies have projected that the behavior of the feedback and software and preferences will vary accordingly, and it causes a mismatch. Very few studies have got solutions for feedback systems for a smartphone-based system [11] and via cross-platform infrastructure for applications related to television[12]

## 2.1 Research Questions

- What factors influence the smart home users' behavior in providing feedback about the smart devices in their houses?
- What are the main features of a user-friendly tool that supports the feedback in innovative technology related to smart devices from the receiver's and senders' perspectives?

## 3. Research Methodology.

This exploratory study used semi-structured interviews to ask a set of group students from the class about gathering feedback on the smart homes. The class was divided into different groups, and the interview should be conducted among the people present in the group. We have used a simple random sampling to select 12 students among 16 students, and it's around 75% of the people where we got the feedback on the smart homes feedback system. We have used three measures to collect the data from the students related to the smart home feedback. We used an open-ended question and an online interview questionnaire in the three measures. We asked users to fill in the answers there and urged users to fill in all the sections and not skip any.

This **expert interview** was conducted on Wednesday, 16th March 2022, with the selected students via **random sampling,** and responses were securely saved on the website. The script related to the interview is included in Appendix A. As users' privacy is an utmost priority, the answers of the users are only visible to the interviewee

## 4. Results:

The data collected from the answers given by the students were downloaded, and preliminary analysis was conducted. The primary objective of this exploratory research is to find the user's attitude towards giving feedback about smart homes. Most users have provided the answer as they have provided feedback to the device after using a particular service. For example, one of the users answered the question "What factors influence users to provide more feedback" as N/A or Null. These answers were eliminated as this could impact the **qualitative analysis**.

## "How often do you fill the feedback forms after using any particular services"?

The qualitative answers entered by the students were not structured. For similar answers, different statements were given, such as "Not very often but I do moderately" and "Not very often, but I sometimes do." These answers have been clubbed as a single response as they both meant the same, which could impact the qualitative analysis approach.

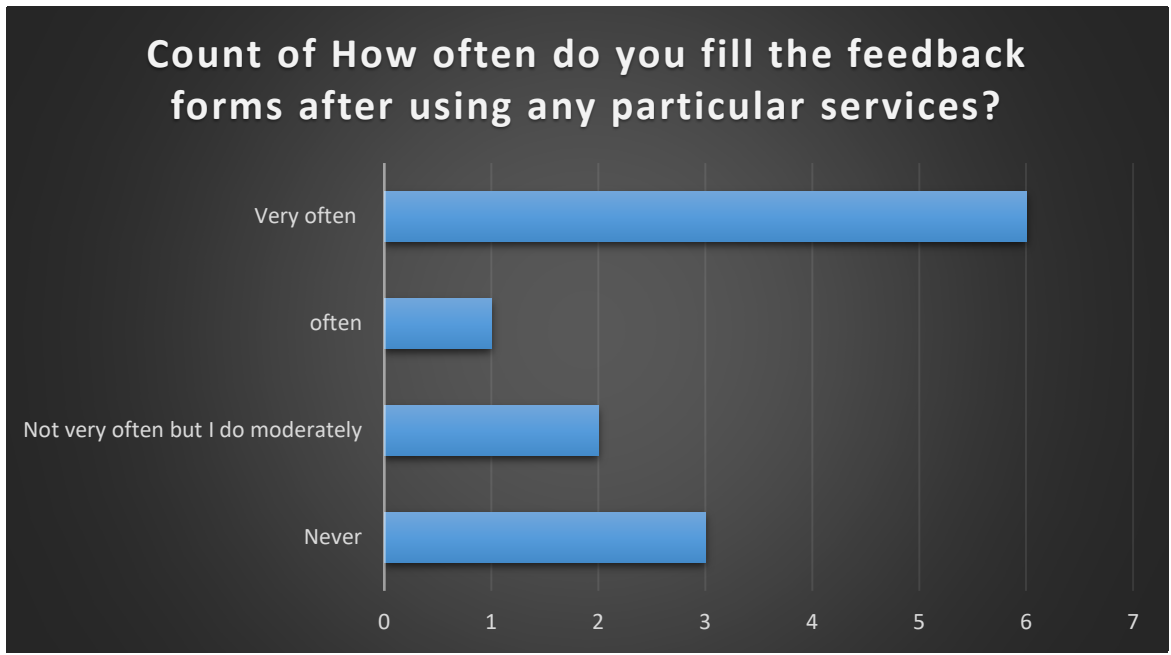

Figure 1: Analysis of the Responses to providing feedback after using services.

Most of the students have answered that they actively provide the feedback after using any services; in our case, it's smart home devices, and the majority of answers provided is "Very Often" as users tend to provide the feedback often. Here it stated that the users liked providing feedback regarding the intelligent homes. However, the second-highest answers show they the users never provided any feedback to the system

## What factors might be influencing users to provide more feedback ratios?

Figure 2 illustrates the qualitative answers provided by the students. The raw answers provided by the students weren't accurate. Hence, the answers are divided into six themes mentioned in the table below.

| THEME OF THE ANSWER | DEFINITION | NO OF STUDENTS |
|---|---|---|
| COUPONS | Providing coupons or offers to fill the feedback | 1 |
| SERVICE | To improve the service and the usability of the product | 6 |
| PERFORMANCE | To improvise the performance of the product | 2 |
| QUALITY | To improvise the device quality | 1 |
| UPGRADES | To provide constant upgrades | 1 |
| UI | To improve the UI of the product | 1 |

**Table 1: Responses given by students.**

The below figure 2 has been projected from the data of Table 1. Around 50% of users have said that the primary factor which impacts after providing the feedback is the "service" of the smart home devices and followed by the "Performance" improvement of the devices. Only one user selected the option called "Quality," as it matters the most.

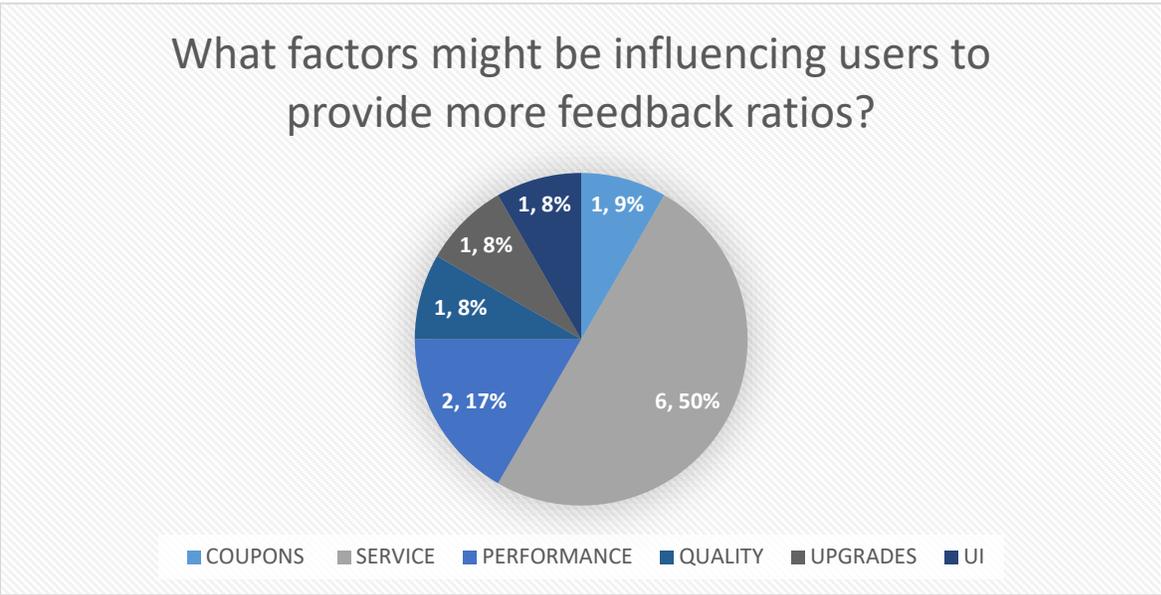

**Figure 2: Analysis of the Responses to providing feedback after using services.**

**Does any mobile helps the audience to provide feedback more satisfactorily and helps in understanding way for both senders and receivers?.**

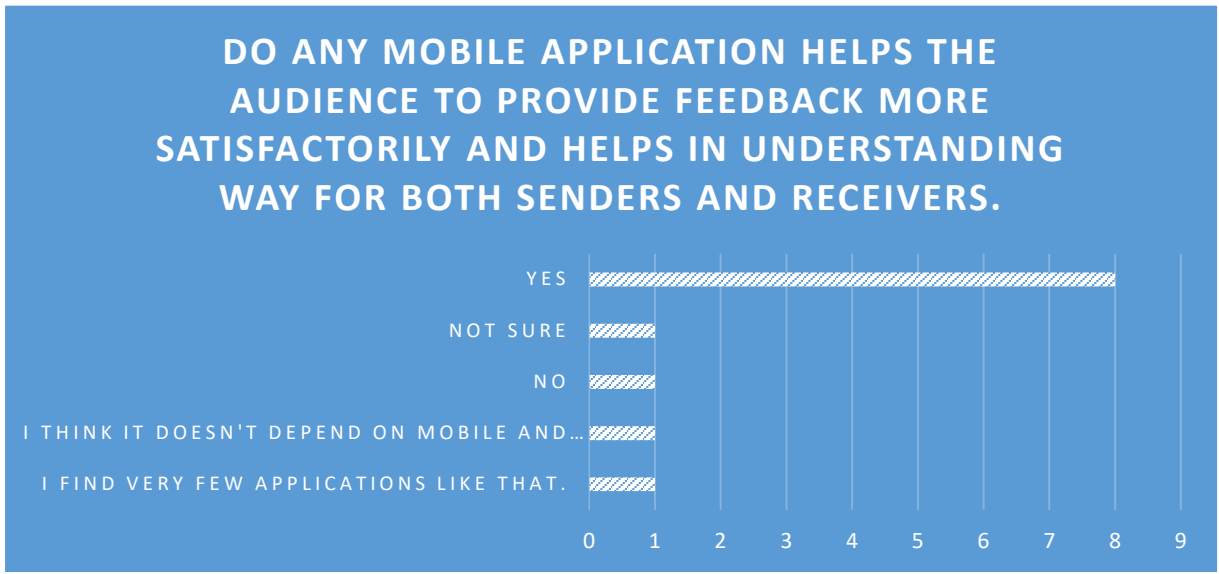

**Figure 3: Analysis of the Responses to providing feedback after using services.**

Figure 3 shows us the users' responses to the question on having any mobile application specifically to provide feedback and help the senders and receivers. Around eight users among 12 have answered as" yes," stating that having such an application helps the user and receiver benefit from the feedback provided. Significantly few users provided various answers such as "No" and "Not Sure" as they think that any application will not help improve the smart homes' feedback system.

The students provided variously typed answers to the questions asked, and when we look at the overview of the answers, the result would be that the users are very eager to provide feedback on the smart home devices, and they want this kind of feedback helps to improvise the product in various ways such as improving the service, performance, UI of the product.

# 5. Discussions:

This exploratory study shows how the users are ready to provide the feedback and how the feedback can be used by users and the. The explorative study with semi-structured interviews has been conducted to know how users feel to give feedback. These answers provided by the students will help to know how eager the users are to provide feedback regarding smart home devices. We talked with understudies to a higher view on the conduct, mentality, and requirements for criticism arrangement on brilliant home gadgets. We accept that the different points of interaction for controlling apparatuses and gadgets and the attributes of savvy home innovations like network, universality, and even intangibility will be tested while practically applying a criticism approach. The users addressed a large portion of the inquiries, which prompted the advancement of a way to deal with the social occasion the input of smart home gadgets.

There is a certain limitation to the data, such as qualitative. The qualitative limitations can be grouped into credibility, transferability, dependability, and confirmability, as suggested by [6]. The answers provided by the users cannot be framed for the analysis. Instead, we have grouped the data into themes or categories and later projected it in a presentable way. Grouping the answers into the themes would be a suggestable way for analysis as it projects what the answer speaks about. The term credibility is very high in the answers given by the students. We have selected around the sample size of 75% of students among the group, which directly impacts the transferability. The random selection of 12 students among 16 students shows the high dependability of the data we received from the students. The validity is shown by taking the answers and providing graphs from the data presented in the results section.

The results align with the findings [2] C.Wilson and Tom Hargreaves, where the results of peer review research on smart home users and the exponential growth of the users. The author projected the functional view, instrumental view of the smart home devices. It also did explain user-related challenges related to smart home challenges and lack of feedback where it matches with how users are very eager to provide the feedback to improvise the smart home devices. After conducting interviews with the users, we have taken sample data, such as how users find it to provide feedback on the devices and the factors impacting the feedback. To make the feedback effects, we even asked users if any mobile application helps in to provide feedback helps users to be more proactive.

We recommend performing even more in-depth studies by exploring and conducting an interview for larger groups on how to provide feedback and how effective the feedback is given will help the users and the receivers to change the design, performance, service, and usability of the smart home devices.

**Appendix A**

IT7010 Information Technology Research Methods
Interview Script

Hello,

Good evening. I hope you're doing well.

My self-santhosh Pogaku. I am part of the research team which conducts an exploratory interview on the Users Approach to Providing Feedback for Smart Home Devices.
This research is anonymous and secure as we don't store any usernames on a form.

Introductory questions
How was your travel to Cincinnati?
Do you use smart home devices at your house?
How often do you use them in your daily life?
Do you follow the latest technology updates in the market?

Interview Questions
How often do you fill out the feedback forms after using any particular services?
What factors might be influencing users to provide more feedback ratios?
Do any mobile helps the audience to provide feedback more satisfactorily and helps in understanding way for both senders and receivers.

Thanks for taking the time and participating in the interview.